\documentstyle[12pt]{article}
\def\hdot{\!\cdot\!}

\begin{document}
\title{Introduction to the Hirota bilinear method} \author{ J. Hietarinta\\
Department of Physics, University of Turku\\ FIN-20014 Turku, Finland\\
e-mail: hietarin@utu.fi}
\date{}
\maketitle

\begin{abstract}
We give an elementary introduction to Hirota's direct method of
constructing multisoliton solutions to integrable nonlinear evolution
equations. We discuss in detail how this works for equations in the
Korteweg--de~Vries class. We also show how Hirota's method can be used
to search for new integrable evolution equations and list the results
that have been obtained before for the mKdV/sG and nlS classes.
\end{abstract}

\section{Why the bilinear form ?}
In 1971 Hirota introduced a new direct method for constructing
multisoliton solutions to integrable nonlinear evolution equations
\cite{Ho71a}. The idea was to make a transformation into new
variables, so that in these new variables multisoliton solutions
appear in a particularly simple form. The method turned out to be very
effective and was quickly shown to give $N$-soliton solutions to the
Korteweg--de~Vries (KdV) \cite{Ho71a}, modified Korteweg--de~Vries
(mKdV) \cite{Ho72a}, sine-Gordon (sG) \cite{Ho72b} and nonlinear
Schr\"odinger (nlS) \cite{Ho73a} equations. It is also useful in
constructing their B\"acklund transformations \cite{Ho74}. Later it
was observed that the new dependent variables (called
``$\tau$-functions'') have very good properties and this has become a
starting point for further developments.

In this talk our aim is to describe how multisoliton solutions can be
constructed using Hirota's method. Multisoliton solutions can, of
course, be derived by many other methods, e.g., by the inverse
scattering transform (IST) and various dressing methods. The advantage
of Hirota's method over the others is that it is algebraic rather than
analytic. The IST method is more powerful (it can handle general
initial conditions) and at the same time more
complicated. Accordingly, if one just wants to find soliton solutions,
Hirota's method is the fastest in producing results.

\section{From nonlinear to bilinear}
The (integrable) PDE's that appear in some particular (physical)
problem are not usually in the best form for further analysis. For
constructing soliton solutions the best form is Hirota's bilinear form
(discussed below) and soliton solutions appear as polynomials of
simple exponentials only in the corresponding new variables. The first
problem we face is therefore to find the bilinearizing transformation.
This is not algorithmic and can sometimes require the introduction of
new dependent and sometimes even independent variables.

Here we will discuss in detail only the KdV equation \begin{equation}
u_{xxx}+6uu_x+u_t=0.
\label{E:KdV}
\end{equation}
Since we have not yet defined what bilinear is, let us first
concentrate on transforming the equation into a form that is quadratic
in the dependent variables. One guideline in searching for the
transformation is that the leading derivative should go together with
the nonlinear term, and, in particular, have the same number of
derivatives. If we count a derivative with respect to $x$ having
degree 1, then to balance the first two terms of (\ref{E:KdV}) $u$
should have degree 2. Thus we introduce the transformation to a new
dependent variable $w$ (having degree 0) by
\begin{equation}
u=\partial_x^2\, w.
\label{E:Bsub}
\end{equation}
After this the KdV equation can be written
\begin{equation}
w_{xxxxx}+6w_{xx}w_{xxx}+w_{xxt}=0,
\end{equation}
which can be integrated once with respect to $x$ to give \begin{equation}
w_{xxxx}+3w_{xx}^2+w_{xt}=0.
\label{E:wKdV}
\end{equation}
In principle this would introduce an integration constant (function of
$t$), but since (\ref{E:Bsub}) defines $w$ only up to $w\to w+x\lambda(t)$,
we can use this freedom to absorb it.

Equations of this form can usually be bilinearized by introducing a new
dependent variable whose natural degree (in the above sense) would be zero,
e.g., $\log F$ or $f/g$. In this case the first one works, so let us define
\begin{equation}
w=\alpha\log F
\end{equation}
with a free parameter $\alpha$. This results in an equation that is fourth
degree in $F$, with the structure \begin{equation}
F^2\times\mbox{(something quadratic)}+
3\alpha (2-\alpha)(2FF''-F'^2)F'^2=0.
\end{equation}
Thus we get a quadratic equation if we choose $\alpha=2$, and the result is
\begin{equation}
F_{xxxx}F-4F_{xxx}F_{x}+3F^2_{xx}+F_{xt}F-F_xF_t=0. \label{E:qKdV}
\end{equation}

In addition to being quadratic in the dependent variables, an equation in
the Hirota bilinear form must also satisfy a condition with respect to the
derivatives: they should only appear in combinations that can be
expressed using Hirota's $D$-operator, which is defined by: \begin{equation}
D_x^n\,f\hdot
g=(\partial_{x_1}-\partial_{x_2})^nf(x_1)g(x_2)\big |_{x_2=x_1=x}.
\end{equation}
Thus $D$ operates on a product of two functions like the Leibnitz rule,
except for a crucial sign difference. For example \begin{eqnarray*}
D_x \,f\hdot g &=& f_{x}g-fg_{x},\\
D_xD_t\, f\hdot g&=&fg_{xt}-f_x g_t - f_tg_x+fg_{xt}. \end{eqnarray*}
Using the $D$-operator we can write (\ref{E:qKdV}) in the following
condensed form
\begin{equation}
(D^4_x+D_xD_t)\,F\hdot F=0.
\label{E:bilKdV}
\end{equation}
To summarize: what we needed in order to obtain the bilinear form
(\ref{E:bilKdV}) for (\ref{E:KdV}) is a dependent variable transformation
\begin{equation}
u=2\partial_x^2 \log F ,
\label{E:subuF}
\end{equation}
and we also had to integrate the equation once.

For a further discussion of bilinearization, see e.g., \cite{HoR,JH:LH}.
Unfortunately the process of bilinearization is far from being algorithmic.
It is even difficult to find out beforehand how many new independent and/or
dependent variables are needed for the bilinearization. In fact for some
equations the natural form may not be bilinear but perhaps trilinear
\cite{multi}. Recently there have been some indications that singularity
analysis can be used to reduce the guesswork: the number of dependent
variables seems to be related to the number of singular manifolds
\cite{Es}.

One important property of equations in Hirota's bilinear form is their
gauge invariance. One can show \cite{multi} that for a quadratic expression
homogeneous in the derivatives, i.e., of the form $\sum_{i=0}^n
c_i\left(\partial_x^i\,f\right) \left(\partial_x^{n-i}\,g\right)$, the
requirement of gauge invariance under $f\to e^{kx}f,\,g\to e^{kx}g$ implies
that the expression can be written in terms of Hirota derivatives. This
gauge invariance can be taken as a starting point for further
generalizations \cite{multi,NEEDS94}.

Finally in this section we would like to list some useful properties of the
bilinear derivative \cite{Ho74}. For $P$ a polynomial,
\begin{eqnarray}
P(D)f\hdot g&=&P(-D)g\hdot f,\\
P(D)1\hdot f=P(-\partial)f,&&P(D)f\hdot1=P(\partial)f,\label{E:Bwith1}\\
P(D)e^{px}\hdot e^{qx}&=&P(p-q)e^{(p+q)x},\\ \partial_x^2 \log
f&=&(D_x^2f\hdot f)/(2f^2),\\ \partial_x^4 \log f&=&(D_x^4f\hdot
f)/(2f^2)-3(D_x^2f\hdot f)^2/(2f^4). \end{eqnarray}

\section{Constructing multisoliton solutions} In this section we will
construct soliton solutions for a class of equations. Properly speaking, the
solutions should be called ``solitary waves'' until we can prove that they
scatter elastically. However, since we are working towards integrable
soliton equations, the solitary wave solutions will at some point become
true solitons, so to make things simple we just call them solitons all the
time.

\subsection{The vacuum, and the one-soliton solution} Now that we have the
KdV equation in the bilinear form, let us start constructing soliton
solutions for it. In fact, it is equally easy to consider a whole class of
bilinear equations of the form \begin{equation}
P(D_x,D_y,...)F\hdot F=0,
\label{E:BK}
\end{equation}
where $P$ is a polynomial in the Hirota partial derivatives $D$. We may
assume that $P$ is even, because the odd terms cancel due to the
antisymmetry of the $D$-operator.

Let us start with the zero-soliton solution or the vacuum. We know that
the KdV equation has a solution $u\equiv 0$ and now we want to find the
corresponding $F$. From (\ref{E:subuF}) we see that
$F=e^{2\phi(t)x+\beta(t)}$ yields a $u$ that solves (\ref{E:KdV}), and
in view of the gauge freedom we can choose $F=1$ as our vacuum
solution. It solves (\ref{E:BK}) provided that \begin{equation}
P(0,0,\dots)=0.  \label{E:0SSc} \end{equation} This is then the first
condition that we have to impose on the polynomial $P$ in (\ref{E:BK}).

The multisoliton solutions are obtained by finite perturbation expansions
around the vacuum $F=1$:
\begin{equation}
F=1+\epsilon \,f_1+\epsilon^2 \,f_2+\epsilon^3 \,f_3+\cdots \end{equation}
Here $\epsilon$ is a formal expansion parameter. For the one-soliton
solution (1SS) only one term is needed. If we substitute \begin{equation}
F=1+\epsilon \,f_1
\label{E:1SS}
\end{equation}
into (\ref{E:BK}) we obtain
\[
P(D_x,\dots)\{ 1\hdot 1 + \epsilon\, 1\hdot f_1 + \epsilon\, f_1\hdot 1 +
\epsilon^2\, f_1\hdot f_1\}=0. \]
The term of order $\epsilon^0$ vanishes because of (\ref{E:0SSc}). For the
terms of order $\epsilon^1$ we use property (\ref{E:Bwith1}) so that, since
now $P$ is even, we get
\begin{equation}
P(\partial_x,\partial_y,\dots)f_1=0.
\label{E:B1}
\end{equation}
The soliton solutions correspond to the exponential solutions of
(\ref{E:B1}). For a 1SS we take an $f_1$ with just one exponential
\begin{equation}
f_1=e^{\eta},\quad \eta=px+qy+\dots+\mbox{ const}, \label{E:1SS1}
\end{equation}
and then (\ref{E:B1}) becomes the {\em dispersion relation} on the
parameters $p,q,\dots$
\begin{equation}
P(p,q,\dots)=0.
\label{E:DR}
\end{equation}
Finally, the order $\epsilon^2$ term vanishes because \[
P(\vec D)e^\eta\hdot e^\eta=e^{2\eta}P(\vec p-\vec p)=0, \]
by (\ref{E:0SSc}).

In summary, the 1SS is given by (\ref{E:1SS},\ref{E:1SS1}) where the
parameters are constrained by (\ref{E:DR}). For KdV the dispersion relation
is $q^3=p$.

\subsection{The two-soliton solution}
The 2SS is built from two 1SS's, and one important principle is that for
integrable systems one must be able to combine {\em any} pair of 1SS's
built on top of the same vacuum. Thus if we have two 1SS's,
$F_1=1+e^{\eta_1}$ and $F_2=1+e^{\eta_2}$, we should be able to combine them
into a form $F=1+f_1+f_2$, where $f_1=e^{\eta_1}+ e^{\eta_1}$. Gauge
invariance suggest that we should try the combination \begin{equation}
F=1+e^{\eta_1}+e^{\eta_2}+A_{12}e^{\eta_1+\eta_2} \end{equation}
where there is just one arbitrary constant $A_{12}$. Substituting this into
(\ref{E:BK}) yields \[
\begin{array}{rccccccl}
P(D)\{\quad1\cdot 1&+&1\cdot e^{\eta_1}&+&1\cdot e^{\eta_2}&+ &
\underline{A_{12}\,1\cdot e^{\eta_1+\eta_2}}& +\\ e^{\eta_1}\cdot
1&+&e^{\eta_1}\cdot
e^{\eta_1}&+&\underline{e^{\eta_1}\cdot e^{\eta_2}}&+
&A_{12}\,e^{\eta_1}\cdot e^{\eta_1+\eta_2}&+\\ e^{\eta_2}\cdot
1&+&\underline{e^{\eta_2}\cdot e^{\eta_1}}&+ &e^{\eta_2}\cdot e^{\eta_2}&+
&A_{12}\,e^{\eta_2}\cdot e^{\eta_1+\eta_2}&+\\
\underline{A_{12}e^{\eta_1+\eta_2}\cdot 1}& +&A_{12}e^{\eta_1+\eta_2}\cdot
e^{\eta_1}&+&A_{12}e^{\eta_1+\eta_2}\cdot e^{\eta_2}&+
&A_{12}^2e^{\eta_1+\eta_2}\cdot e^{\eta_1+\eta_2}&\quad\}=0. \end{array}
\]
In this equation all non-underlined terms vanish due to
(\ref{E:0SSc},\ref{E:DR}). Since $P$ is even, the underlined terms combine
as $
2A_{12}P(\vec p_1+\vec p_2)+2P(\vec p_1-\vec p_2)=0,$ from which $A_{12}$
can be solved as
\begin{equation}
A_{12}=-\frac{P(\vec p_1-\vec p_2)}{P(\vec p_1+\vec p_2)}. \label{E:pf}
\end{equation}

The important thing about this result is that we were able to construct a
two-soliton solution for a huge class of equations, namely all those whose
bilinear form is of type (\ref{E:BK}). In particular this includes many
non-integrable systems.

\subsection{Multi-soliton solutions}
The above shows that for the KdV class (\ref{E:BK}) the existence of 2SS is
not strongly related to integrability, but it turns out that the existence
on 3SS is very restrictive.

A 3SS should start with $f_1=e^{\eta_1}+e^{\eta_2}+e^{\eta_3}$ and, if the
above is any guide, contain terms up to $f_3$. If we now use the
requirement that the solution should reduce to a 2SS when the third soliton
goes to infinity (which corresponds to $\eta_k\to\pm\infty$) then one finds
that $F$ must have the form \begin{eqnarray}
F&=&1+e^{\eta_1}+e^{\eta_2}+e^{\eta_3}\nonumber\\ &&\quad
+A_{12}e^{\eta_1+\eta_2}+A_{13}e^{\eta_1+\eta_3}+
A_{23}e^{\eta_2+\eta_3}+A_{12}A_{13}A_{23}e^{\eta_1+\eta_2+\eta_3}.
\label{E:3SS}
\end{eqnarray}
Note in particular that this expression contains no additional freedom. The
parameters $p_i$ are only required to satisfy the dispersion relation
(\ref{E:DR}) and the phase factors $A$ were already determined
(\ref{E:pf}). This extends to NSS \cite{HoKK}: \begin{equation}
F=\sum_{\mu_i=0,1 \atop 1\le i\le N}\,\exp\left(\sum_{1\le i<j\le N}
\varphi(i,j)\mu_i\mu_j+\sum_{i=1}^N\mu_i\eta_i\right) , \end{equation}
where $(A_{ij}=e^{\varphi(i,j)})$. Thus the ansatz for a NSS is completely
fixed and the requirement that it be a solution of (\ref{E:BK}) implies
conditions on the equation itself. Only for integrable equations can we
combine solitons in this simple way. More precisely, let us make the

\vskip 0.3cm
\noindent
{\sc Definition:} {\it A set of equations written in the Hirota bilinear
form is {\bf Hirota integrable}, if one can combine any number $N$ of
one-soliton solutions into an NSS, and the combination is a finite
polynomial in the $e^\eta$'s involved.} \vskip 0.3cm

\noindent
In all cases known so far, Hirota integrability has turned out to be
equivalent to more conventional definitions of integrability.

\section{Searching for integrable evolution equations.} Since the existence
of a 3SS is very restrictive, one can use it as a method for searching for
new integrable equations. All search methods depend on some initial
assumptions about the structure. In this case we assume that the nonlinear
PDE can be put into a bilinear form of type (\ref{E:BK}), but no
assumptions are made for example on the number of independent variables.

\subsection{KdV}
If one now substitutes (\ref{E:3SS}) into (\ref{E:BK}) one obtains the
condition
\begin{eqnarray}
&&\sum_{\sigma_i=\pm 1}P(\sigma_1\vec p_1+\sigma_2\vec p_2+\sigma_3\vec
p_3) \nonumber\\
&&\ \qquad \times P(\sigma_1\vec p_1-\sigma_2\vec p_2) P(\sigma_2\vec
p_2-\sigma_3\vec p_3)
P(\sigma_3\vec p_3-\sigma_1\vec p_1)\stackrel{.}{=}0, \label{E:3SC}
\end{eqnarray}
where the symbol $\stackrel{.}{=}0$ means that the equation is required to
hold only when the parameters $\vec p_i$ satisfy the dispersion relation
$P(\vec p_i)=0$.

In order to find possible solutions of (\ref{E:3SC}) we made a computer
assisted study \cite{JH:S1} and the result was that the only genuinely
nonlinear equations that solved (\ref{E:3SC}) were \begin{eqnarray}
(D_x^4-4D_xD_t+3D_y^2)F\cdot F &=&0,\label{E:KP}\\
(D_x^3D_t+aD_x^2+D_tD_y)F\cdot F &=&0,\label{E:HSI}\\
(D_x^4-D_xD_t^3+aD_x^2+bD_xD_t+cD_t^2)F\cdot F &=&0,\label{E:my}\\
(D_x^6+5D_x^3D_t-5D_t^2+D_xDy)F\cdot F &=&0.\label{E:SKR} \end{eqnarray}
and their reductions. These equations also have 4SS and they all pass the
Painlev\'e test \cite{PT}. Among them we recognize the 
Kadomtsev-Petviashvili (containing
KdV and Boussinesq) (\ref{E:KP}), Hirota-Satsuma-Ito (\ref{E:HSI}) and
Sawada-Kotera-Ramani (\ref{E:SKR}) equations; they appear in the Jimbo-Miwa
classification \cite{JM}. The only new equation is (\ref{E:my}). It is
somewhat mysterious. It has not been identified within the Jimbo-Miwa
classification because it has no nontrivial scaling invariances,
furthermore we do not know its Lax pair or B\"acklund transformation. But
it does have at least 4SS, and it passes the Painlev\'e test \cite{PT}.

\subsection{mKdV and sG}
As was mentioned before, Hirota's bilinear method has been applied to many
other equations beside KdV. Here we would like to mention briefly some of
them.

For example the modified Korteweg--de Vries (mKdV) and sine--Gordon (sG)
equations have a bilinear form of the type \begin{equation}
\left\{\begin{array}{r}
B(D_{\vec x})\, G\cdot F=0,\\ A(D_{\vec x})(F\cdot F+G\cdot G) =0,
\end{array}\right.
\label{E:mKdVt}
\end{equation}
where $A$ is even and $B$ either odd (mKdV) or even (sG). For mKdV we have
$B=D_x^3+D_t,\,A=D_x^2$, and for SG, $B=D_xD_y-1,\,A=D_xD_y$. This class of
equations also has 2SS for any choice of $A$ and $B$. If $B$ is odd one can
make a rotation $F=f+g,\,G=i(f-g)$ after which the pair (\ref{E:mKdVt})
becomes $B\, g\cdot f=0,\,A \, g\cdot f=0$.

In principle the pair (\ref{E:mKdVt}) can have two different kinds of 
solitons,
\begin{equation}
\left\{\begin{array}{ll}
F=1+e^{\eta_A},\, G=0,&\mbox{with dispersion relation } A(\vec p)=0,\\
F=1,\, G=e^{\eta_B},&\mbox{with dispersion relation } B(\vec p)=0.
\end{array}\right.
\end{equation}
In mKdV and SG the $A$ polynomial is too trivial to make the first kind of
soliton interesting. In \cite{JH:S2,JH:LH} we searched for polynomials $A$
and $B$ for which any set of three solitons could be combined for a 3SS.
The final result contains 5 equation, of mKdV type, three of them have a
nonlinear $B$ polynomial but a factorizable $A$ part (and hence only one
kind of soliton with $B$ acting as the dispersion relation),
\begin{eqnarray}
\left\{\begin{array}{rcl}
(aD_x^7+bD_x^5+D_x^2D_t+D_y)\, G\cdot F&=&0,\\ D_x^2\, G\cdot F&=&0,
\end{array}\right.
\\
\left\{\begin{array}{rcl}
(aD_x^3+bD_x^3+D_y)\, G\cdot F&=&0,\\
D_xD_t\, G\cdot F&=&0,
\end{array}\right.
\\
\left\{\begin{array}{rcl}
(D_xD_yD_t+aD_x+bD_t)\, G\cdot F&=&0,\\
D_xD_t\, G\cdot F&=&0.
\end{array}\right.
\end{eqnarray}
For a discussion of the nonlinear versions of the last two equations, see
\cite{Nimmo}.

In two cases both $A$ and $B$ are nonlinear enough to support solitons,
note that the $B$ polynomials are the same and that the $A$ parts have
already appeared in the KdV list: \begin{eqnarray}
\left\{\begin{array}{rcl}
(D_x^3+D_y)\, G\cdot F&=&0,\\
(D_x^3D_t+aD_x^2+D_tD_y)\, G\cdot F&=&0, \end{array}\right.
\\
\left\{\begin{array}{rcl}
(D_x^3+D_y)\, G\cdot F&=&0,\\
(D_x^6+5D_x^3D_y-5D_y^2+D_tD_x)\, G\cdot F&=&0. \end{array}\right.
\end{eqnarray}

Two equations of sine-Gordon type were also found: \begin{eqnarray}
\left\{\begin{array}{rcl}
(D_xD_t+b)\, G\cdot F&=&0,\\
(D_x^3D_t+3bD_x^2+D_tD_y)(F\cdot F+G\cdot G)&=&0, \end{array}\right.
\\
\left\{\begin{array}{rcl}
(aD_x^3D_t+D_tD_y+b)\, G\cdot F&=&0,\\
D_xD_t(F\cdot F+G\cdot G)&=&0.
\end{array}\right.
\end{eqnarray}

\subsection{nlS}
A similar search was performed \cite{JH:S4,JH:LH,JHnls} on equations of
nonlinear Schr\"odinger (nlS) type,
\begin{equation}
\left\{\begin{array}{rcl}
B(D_{\vec x})\, G\cdot F&=&0,\\
A(D_{\vec x}) \, F\cdot F&=&|G|^2,
\end{array}\right.
\end{equation}
where $F$ is real and $G$ complex. Again two kinds of solitons exist,
\begin{equation}
\left\{\begin{array}{ll}
F=1+e^{\eta_A},\, G=0,&\mbox{with dispersion relation } A(\vec p)=0,\\
F=1+Ke^{\eta_B+\eta_B^*},\, G=e^{\eta_B},& \mbox{with dispersion relation }
B(\vec p)=0. \end{array}\right.
\end{equation}
In this case the existence of a 2SS is not automatic because the
1SS already involves terms of order $\epsilon^2$ and a 2SS therefore
$\epsilon^4$, whereas in the previous cases $\epsilon^2$ contributions were
sufficient for 2SS.

Three equations were found in this search: \begin{eqnarray}
\left\{\begin{array}{rcl}
(D_x^2+iD_y+c)\, G\cdot F&=&0,\\
(a(D_x^4-3D_y^2)+D_xD_t)\, F\cdot F&=&|G|^2, \end{array}\right.
\\
\left\{\begin{array}{rcl}
(i\alpha D_x^3+3cD_x^2+i(bD_x-2dD_t)+g)\, G\cdot F&=&0,\\ (\alpha
D_x^3D_t+aD_x^2+(b+3c^2)D_xD_t+dD_t^2)\, F\cdot F&=&|G|^2,
\end{array}\right.
\\
\left\{\begin{array}{rcl}
(i\alpha D_x^3+3D_xD_y-2iD_t+c)\, G\cdot F&=&0,\\
(a(\alpha^2D_x^4-3D_y^2+4\alpha D_xD_t)+bD_x^2)\, F\cdot F&=&|G|^2.
\end{array}\right.
\label{E:kpds}
\end{eqnarray}
Perhaps the most interesting new equation above is the combination in
(\ref{E:kpds}) of the two most important $(2+1)$-dimensional equations,
Davey-Stewartson and Kadomtsev-Petviashvili equations, see \cite{JHnls},
page 315.

The above lists contains many equations of which nothing else is known
other than that they have 3SS and 4SS. This alone suggests that they are
good candidates for integrable $2+1$ dimensional equations, but connections
to other definitions of integrability (like Lax pairs) are still open.

\section*{Acknowledgments}
This work was supported in part by the Academy of Finland, project 31445.

\end{document}